\documentclass[prl, twocolumn,preprintnumbers,amsmath,amssymb]{revtex4}
\usepackage{graphicx}
\usepackage{dcolumn}
\usepackage{bm}
\usepackage{color}

\begin{document}

 \newcommand{\com}[1]{\textcolor{red}{(#1)}}

\title{  Accurate $\emph{in situ}$ Measurement of  Ellipticity  Based on Sub-cycle Ionization Dynamics}

\author{Chuncheng Wang,$^{1}$\footnote{These authors equally contribute  to this work.}}
\author{Xiaokai Li$^{1*}$}
\author{Xiang-Ru Xiao$^{2*}$}
\author{Yizhang Yang$^{1}$}
\author{Sizuo Luo$^{1}$}
\author{Xitao Yu$^{1}$}
\author{Xinpeng Xu$^{1}$}
\author{Liang-You Peng$^{2,3}$}
\email{liangyou.peng@pku.edu.cn}
\author{Qihuang Gong$^{2,3}$}
\author{Dajun Ding$^{1}$}
\email{dajund@jlu.edu.cn}

\affiliation{$^{1}$Institute of Atomic and Molecular Physics,
Jilin University, Changchun 130012, China}

\affiliation{$^{2}$State Key Laboratory for Mesoscopic Physics  and  Collaborative Innovation Center of Quantum Matter, School of Physics,  Peking University, Beijing 100871, China}

\affiliation{$^{3}$Collaborative Innovation Center of Extreme Optics, Shanxi University, Taiyuan, Shanxi 030006, China}

\date{\today}

\begin{abstract}

 Elliptically polarized laser pulses  (EPLPs) are widely applied in many fields of ultrafast sciences, but
 the  ellipticity~($\varepsilon$)  has never been $\emph{in situ}$ measured in the interaction zone of the laser focus. In this work, we propose and realize a robust scheme to retrieve the  $\varepsilon$ by temporally overlapping two identical counter-rotating EPLPs.  The combined linearly electric field is  coherently controlled to ionize Xe atoms by varying the phase delay between the two EPLPs. The electron spectra of the above-threshold ionization and the ion yield are sensitively modulated by the phase delay. We demonstrate that these modulations can be used to accurately determine  $\varepsilon$ of  the EPLP. We show that the present method is highly reliable and  is applicable in a wide range of laser parameters.  The accurate retrieval of  $\varepsilon$ offers a better characterization of a laser pulse, promising a more  delicate and quantitative  control of the sub-cycle dynamics in many strong field processes.

\end{abstract}

\maketitle

It is well known that many of the strong-field phenomena are extremely sensitive to the electric waveform of the laser pulse, as their first step is the electronic tunneling whose  rate exponentially depends  on the instantaneous field strength~\cite{Landau}. In order to coherently control the ultrafast dynamics, enormous efforts have been made to manipulate various kinds of laser parameters or to synthesize a desirable shaped-pulse using a two-color~\cite{twocl1, twocl2} or a multiple-color~\cite{Krausz_Syn} field. In doing so, one has to precisely characterize or measure the essential quantities of a laser pulse, including the  ellipticity~($\varepsilon$) for an elliptically polarized light.

Due to its suppression of the electron recollision~\cite{Corkum1993} in some extent,  the importance of the ellipticity  has been demonstrated in many different processes, such as the strong field single ionization~\cite{keller2013, atto3, dorner1}, the molecular chirality~\cite{chiral2},  the double ionization~\cite{release1,LiWen,angular1, Ma2018}, the high harmonic generation~(HHG)~\cite{HHG1}, and the ring current~\cite{dorner2}.   A pair of single-color counter-rotating circularly polarized pulses are used in the polarization gating method to generate a single attosecond pulse~\cite{ polargate2,polargate3}, which, by adding a weak second harmonic field, has been developed to the double optical gating~(DOG)~\cite{DOG1} and later to the generalized DOG~(GDOG)\cite{GDOG2, DOG2} scheme. In addition,  schemes of two-color counter-rotating field  have found important  applications in many topics such as the double ionization~\cite{CRTC3, CRTC1,CRTC2}, the laser-induced electron diffraction~\cite{Blaga2012, Wang2012},  the spin polarization~\cite{CRTC6, CRTC5}, and the generation of extreme ultraviolet lights~\cite{Fleischer,denitsa}.

 Apparently, the ellipticity $\varepsilon$ of  an elliptically polarized laser pulse~(EPLP) is one of the most important quantities,  which has played versatile roles in controlling the strong field dynamics and in their various applications. However, according to the best of our knowledge, the accurate value of   $\varepsilon$  in the interaction zone has never been measured $\emph{in situ}$. It is well known that a perfect circularly polarization is unrealistic after focusing in the vacuum due to the imperfect optics performance for ultrashort laser pulses.  In most laboratories, the $\varepsilon$ is usually roughly estimated by monitoring the transmitted pulse energy after the polarizer, which severely loses its accuracy for a high $\varepsilon$ value. Therefore, researchers usually assume that one has a circularly polarized pulse when the estimated $\varepsilon>0.9$. However,  it is critical to know the accurate value of $\varepsilon$ in many experiments, e.g.,  the streaking angle  related to the time delay strongly depends  on $\varepsilon$ in the attoclock-type measurements~\cite{attoclock1,attoclock2,attoclock3}.

 In this Letter,  we establish an $\emph{in situ}$ experimental method for measuring the accurate ellipticity  of a laser pulse, which is based on the widely available setup for the interaction of the counter-rotating EPLPs with rare gas atoms.  This scheme provides a delicate  electric field variation  with a well-controlled polarization rotation. The maximum of angular distribution of electrons in  the above-threshold ionization~(ATI) synchronizes with the  polarization rotation,  which directly tags the phase delay between the two pulses. The observed phase dependent Stark-shift of ATI electrons and the ionization yield oscillations  reflect exactly the minimum/maximum electric field ratio of the counter-rotating field within one optical cycle,  which provides us with an accurate value of $\varepsilon$ for the EPLPs.  We show that our scheme accurately determines a high ellipticity of 0.963 with an uncertainty of 0.007.   The measurements based on our proposal are quantitatively confirmed by our theories based on the numerical solution to the time-dependent Schr\"odinger equation~(TDSE)~\cite{tdse1,tdse2} and by the Perelomov-Popov-Terent'ev~(PPT) theory~\cite{Perelomo1,Perelomo2,Perelomo3,ppt1,ppt2}.
We show that our method provides an easy and transferable calibration standard of the ellipticity  for laser pulses covering a wide range of laser parameters.

\begin{figure}
\begin{center}
\includegraphics[width=3.5in]{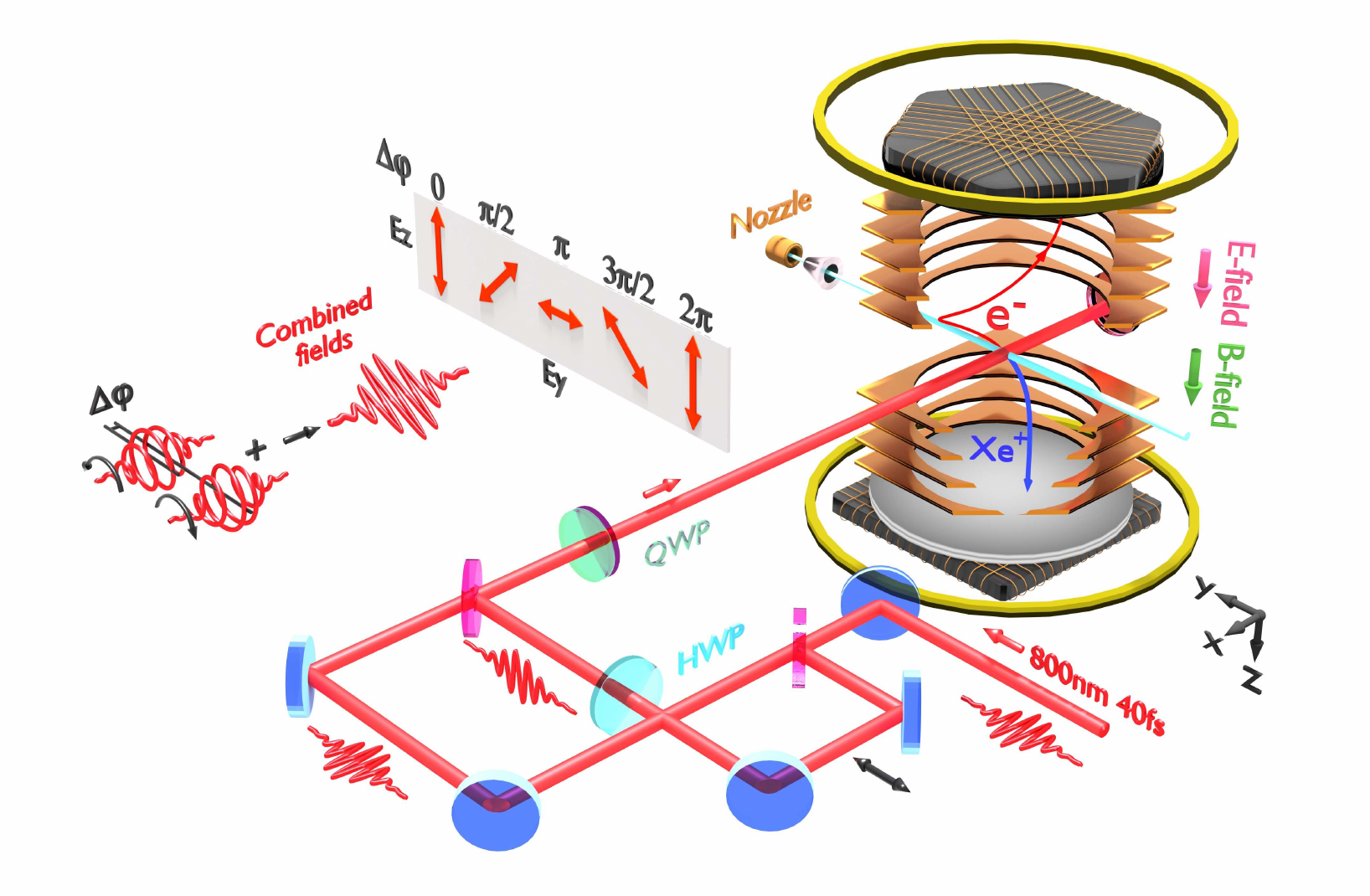}
\end{center}
\caption{(Color online)
Schematic diagram of the experimental setup. The laser propagates along the $x$ axis and the polarization plane of each near-circular pulse lies in the $y$-$z$ plane. The insert  shows the temporal overlap
between the two counter-rotating pulses and  the combined linearly polarized field with the rotation of its polarization  and  variation in the peak field strength  for   different phase delay $\Delta \varphi$.}
\label{fig1}
\end{figure}

The experimental setup~\cite{setup} is depicted in Fig.~1, in which a linearly polarized 800~nm, 40 femtosecond laser pulse is divided into two arms using a 50$\%$ dielectric beam splitter.  One arm is delayed by the translation stage and its polarization is rotated to the $z$ axis by a half-wave plate~(HWP).  The polarization of the other arm  is kept in the $y$ axis. {Then the two beams with orthogonal polarizations are converted to the counter-rotating near-circularly laser pulses by a quarter-wave plate~(QWP).The pulse energy and ellipticity of each arm are kept the same during the measurement. When the time delay between the two beams is limited in a small range, the combination of the two elliptical pulses is equivalent to a linearly polarized field, as shown in the insert of Fig.~1. The electric field amplitude of the counter-rotating EPLPs can be expressed as  $F_0=2F_{\text{max}}\sqrt{\cos^2\frac{\Delta\varphi}{2}+\varepsilon^{2} \sin^2\frac{\Delta\varphi}{2}}$ where $F_{\text{max}}$ is the electric field along the major axis of the elliptical pulse and $\Delta\varphi$ is the phase delay between the two pulses. The direction of the polarization $\theta_L$  satisfies $\tan\theta_L=-\varepsilon\tan\frac{\Delta\varphi}{2}$. By increasing the time delay between the two beams, the polarization of the counter-rotating field rotates and the amplitude of the field varies with a period of 2$\pi$.}  By the traditional optical method through monitoring the transmitted pulse energy after the polarizer,  the highest  $\varepsilon$ we can measure is  0.93 for each elliptical arm.  Then, the {counter-rotating EPLPs} go through a  quartz window of a thickness about 1 mm and {are} then focused by a concave mirror with a  focal length of 75~mm. A xenon gas target is introduced by the supersonic expansion with a constant gas pressure to interact with the counter-rotating field.  The  COLTRIMS~(cold-target recoil-ion momentum spectroscopy) is used to detect the produced ions and electrons~\cite{setup2}. The three-dimensional momentum vector of a charged particle is extracted from the measured time of flight and the position of each particle.

The energy-integrated angular distributions of ATI electrons at different time delays determined by $\Delta\varphi$ are shown in Fig.~2(a). The maximum ejection direction $\theta_m$ of the electrons    gradually  shifts from $ -45^\circ$   to $-180^\circ$    when  $\Delta\varphi$ changes from  $0.5\pi$ to  $2\pi$. Since the ATI electrons will be dominantly emitted  along the laser polarization direction of the { counter-rotating field}, the variation of $\theta_m$ directly reflects the polarization rotation of the { combined counter-rotating field} at different time delays, i.e. $\theta_m \approx \theta_L$. Please note that $  \theta_L \approx - \Delta\varphi/2$  for a  laser with a high ellipticity  $\varepsilon$. From Fig.~2(a), one observes that the measured angular distributions~(symbols) agree quite well with  those of the theoretical calculations~(solid lines),  which are based on the numerical solution to the TDSE.  An effective model potential \cite{model_potential} is used in our TDSE calculations, and all the shown results are the average of the ones starting from the initial $\rm{p}_0$, $\rm{p}_+$ and $\rm{p}_-$ states.    This  agreement allows us to $\emph{in situ}$ tag the phase delay between the two laser pulses and can guarantee the phase-delay stability within a minimal deviation from the experimental fluctuations.

\begin{figure}
	\begin{center}
		\includegraphics[width=3.2in]{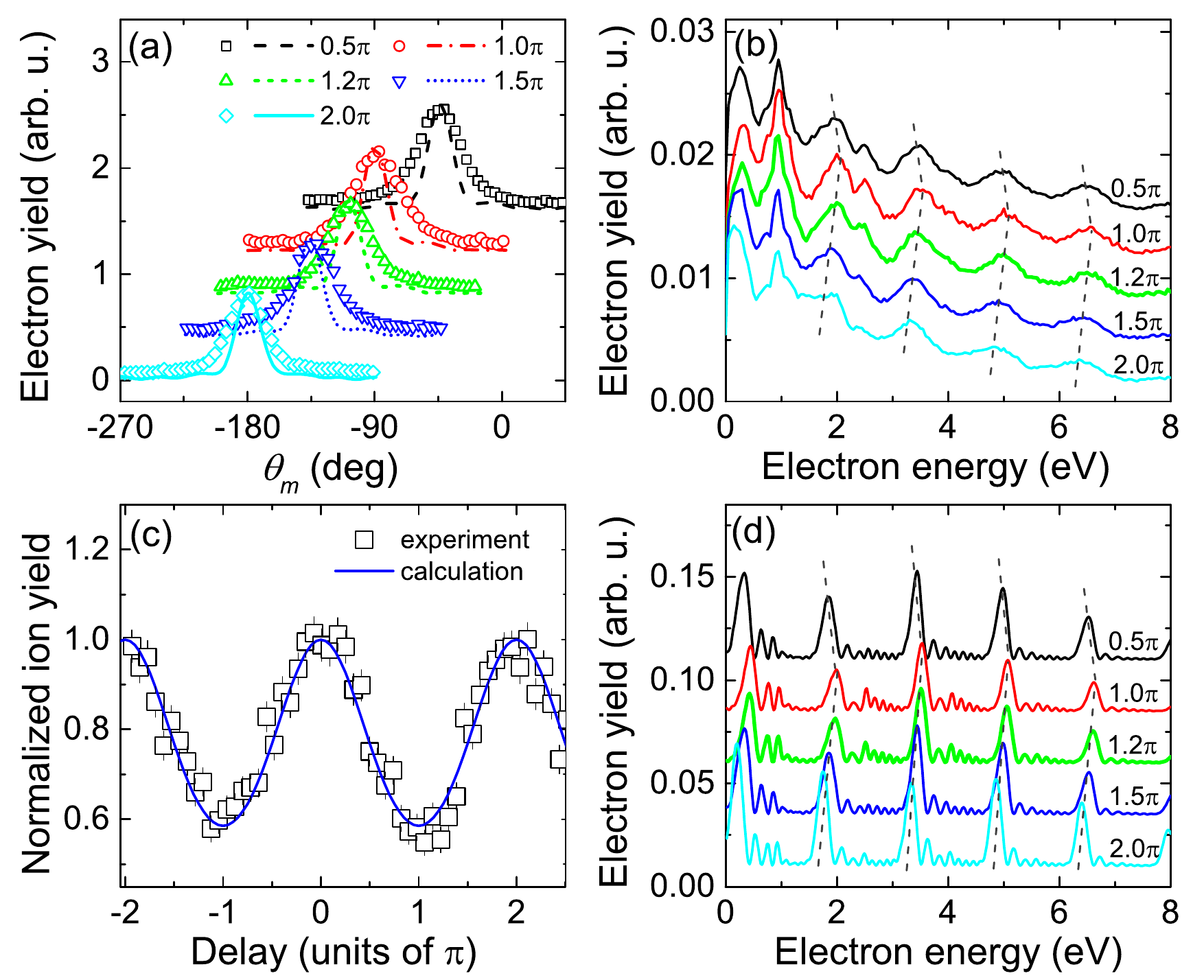}
	\end{center}
	\caption{(Color online) The phase-delay dependence of the ATI electrons and the ion yield. (a) The energy-integrated angular distributions at various delays~(symbols for the measurements and lines for the TDSE simulations). (b) The experimentally measured angle-integrated ATI spectra at different delays, compared against the TDSE calculations under the same laser parameters shown in (d). In  (a),(b), and (d), the curve for each delay is relatively shifted vertically for a better visibility.  In (c), the normalized ion yield for the single ionization of Xe is shown at different phase delays from the measurements~(squares) and the PPT calculations~(solid line). }
	\label{fig2}
\end{figure}

In order to reveal the dependence of the ATI spectrum on the actual  peak intensity of the {counter-rotating field}, in Fig.~2(b),  we present  the angle-integrated ATI spectra measured at those various time delays. One observes clean and well-separated ATI peaks with a tiny but clear shift~(guided by dashed lines)  when the phase delay $\Delta\varphi$ is   gradually changed.   Actually, the peak shift exhibits a clear phase-dependent oscillation with a period of   $2\pi$.  For the nonresonant ionization, it is known that the ATI peaks  can be described as~\cite{ati_up}:
$
E = n \omega -(I_{p}+ U_{p}),
\label{ATI}
$
 where $n$ is the total number of absorbed photons, $\omega$ is the laser   frequency, $I_{p}$ is the ionization potential, and $U_{p}=\frac{e^{2}F_0^{2}}{4m_e\omega^{2}}$ is the ponderomotive energy. The ATI peaks may shift for different laser intensities since the ionization threshold has an ac Stark-shift that is equal to $U_{p}$.  The observed  oscillation of the tiny peak shift  comes from the change of $U_{p}$, which is proportional to the actual peak laser intensity of the counter-rotating field.  As mentioned previously, the strength of the { counter-rotating field} oscillates with the phase delay $\Delta\varphi$,  and its minimum and maximum  appear at  $\pi$ and  $2\pi$ respectively, due to the constructive interference of the minor and major axis for the two counter-rotating EPLPs.

 Therefore, the observed peak shift oscillation exactly reflects the behavior of the counter-rotating field. In particular, the minimum/maximum electric field ratio of the combined counter-rotating field directly relates to the field ratio~($\varepsilon$) along the minor and major axis of  the EPLP.  By accurately calibrating the peak intensity of the counter-rotating field, one can extract an $\emph{in situ }$ value of $\varepsilon$ for  the EPLP.  The peak laser intensity can first be roughly calibrated  by measuring the ionized electron drift momentum with one arm of the near-circular laser pulse~\cite{diftmomentum}, and then an accurate intensity can be evaluated from the observed $U_{p}$ shift in the ATI spectra at  different phase delays.  By doing so,  we obtain a  minimum intensity  of  47.5~TW/cm$^{2}$ at $\pi$ and  a maximum intensity of 51.2~TW/cm$^{2}$ at   $2\pi$. We find that the intensities extracted from different orders of ATI peaks show excellent agreement, which   only gives a statistic deviation of about $1.5\%$.    With this accurate laser intensity,  an $\emph{in situ }$ $\varepsilon$ of $0.963\pm0.007$ for  the EPLP can be directly derived from the ratio of intensities  at phase delay $\pi$ and $2\pi$.  The uncertainty 0.007 mainly propagates from the intensity calibration.  Please note that, by carefully adjusting the optics, one can make sure the two arms of the EPLPs are almost identical. Of course, the imperfectness of optical components may induce a slight difference in the ellipticities of the two pulses~(less than 0.003 for our experiment). In this case, the $\emph{in situ }$ extracted  $\varepsilon$ is actually the average ellipticity of two counter-rotating EPLPs, which however can unambiguously represent  the ellipticity of each EPLP with a sufficient accuracy~\cite{sup}. In addition, we point out that, by simply blocking one of the   EPLPs,  one can directly apply the other well calibrated EPLP  in the further researches of ultrafast sciences.

 To confirm the accuracy of the measurement, we carry out a series of TDSE calculations for the ATI spectra, with the same laser parameters~(intensities and $\varepsilon$) extracted from the experimental measurements. The angle-integrated ATI spectra from TDSE are shown in Fig.~2(d) for various  phase delay between the two laser pulses.   As can be seen, these theoretical results indeed perfectly agree  with those experimental measurements shown in Fig.~2(b). As a consistent cross-check, one can also extract the value of $\varepsilon$ from the theoretically calculated ATI spectra, using the same procedures  as those for the experiment. We get a theoretical value of $\varepsilon$  to be 0.964,  which retracts the input experimental measurement of 0.963 within a deviation of 0.001.  The above comparison  fully validates   our methodology and also the accuracy of the experimental measurement of $\varepsilon$. Please note that, the $\emph{in situ }$  value $\varepsilon$ is  shown to be larger than the estimated value of 0.93 by the traditional optical method, indicating that the ultrafast pulses become more isotropic in the focus region.  This difference may originate from either the inaccurate optical measurement at a large ellipticity, or  the imperfect performance of the window and reflection  mirror. One notes that similar ATI oscillations can also be observed  by simply inserting a rotating HWP and a polarizer after the single EPLP, which allows the extraction of  $\varepsilon$ using the same method.  However, this approach lacks of accuracy for near-circularly polarized laser pulses and also relies on the parameters of polarizer.

\begin{figure}
\begin{center}
\includegraphics[width=3.5in]{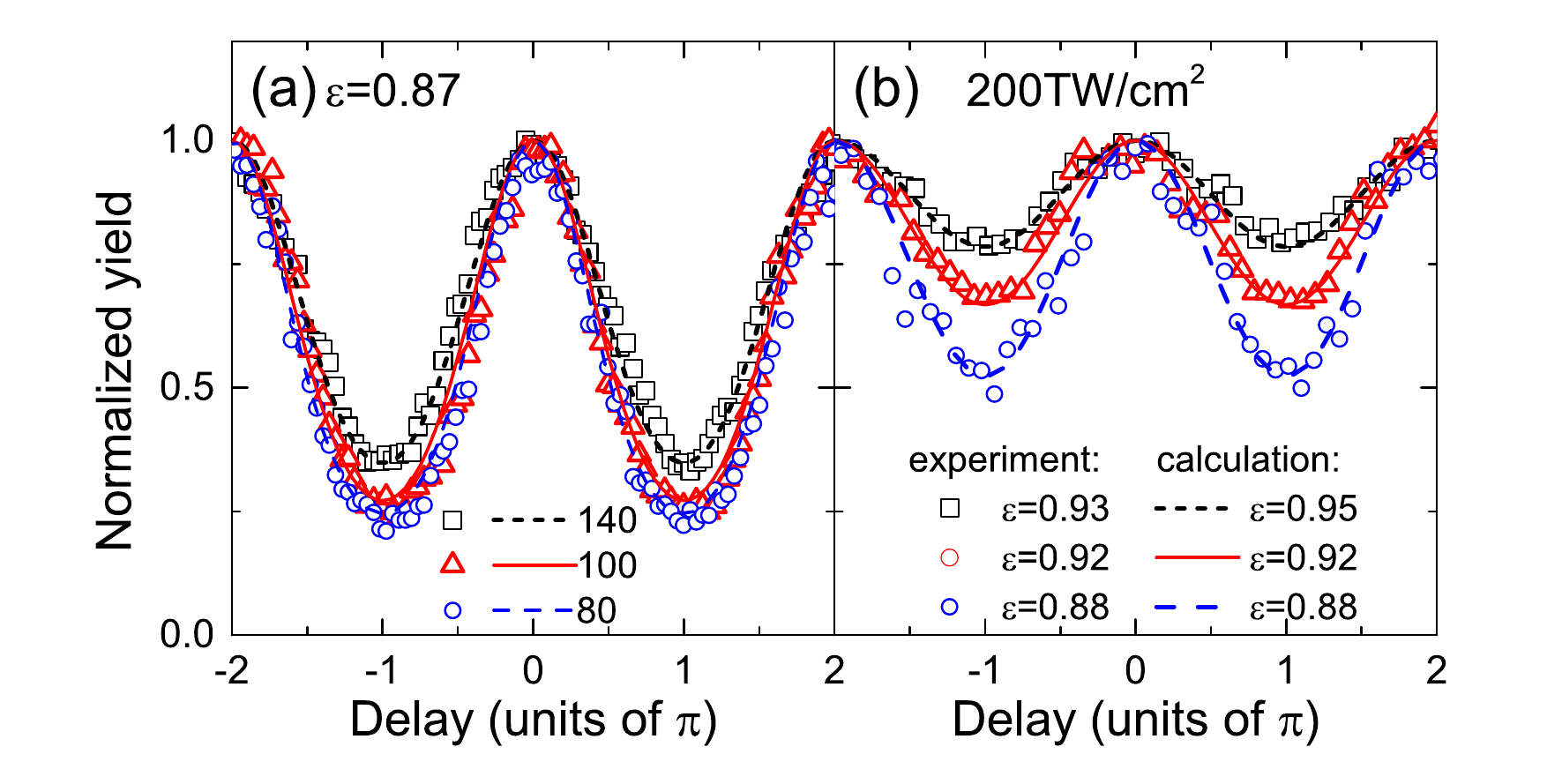}
\end{center}
\caption{(Color online) The normalized ion yield  as a function of the phase delay for (a) different laser intensities~(in TW/cm$ ^{2}$) at a fixed  $\varepsilon =0.87$; (b) various values of the ellipticity at a fixed laser intensity of 200~TW/cm$ ^{2}$. For every case, the lines stand for the theoretical calculations, while the symbols for the experimental measurements where the value of the ellipticity is estimated by the usual optical method. }
\label{fig3}
\end{figure}

Apart from the  peak shift oscillation of the  ATI spectra, the measured total ion yield of the single ionization will be also periodically modulated by the phase delay of the two pulses,  as shown in Fig.~2(c) by the squares.  Theoretically, the total ion yield can be simply calculated by the PPT theory for the combined linearly polarized field  for the  $\rm{p}_0$ orbital with the focussing volume effects fully taken into account~\cite{ppt1,ppt2}. In the simulations, we  use the same laser intensities calibrated by the above ATI measurement, but  slightly adjust the particular value of $\varepsilon$ to best fit the modulation of the experimental  measurement.    When   we take $\varepsilon =0.957$ in the simulation,   the two  normalized curves perfectly agree with each other, which means that the accuracy of $\varepsilon$ extracted from the ion yield oscillation measurement is  better than 0.01, compared to the experimental value of 0.963.

\begin{figure}[t]
\begin{center}
\includegraphics[width=3.2in]{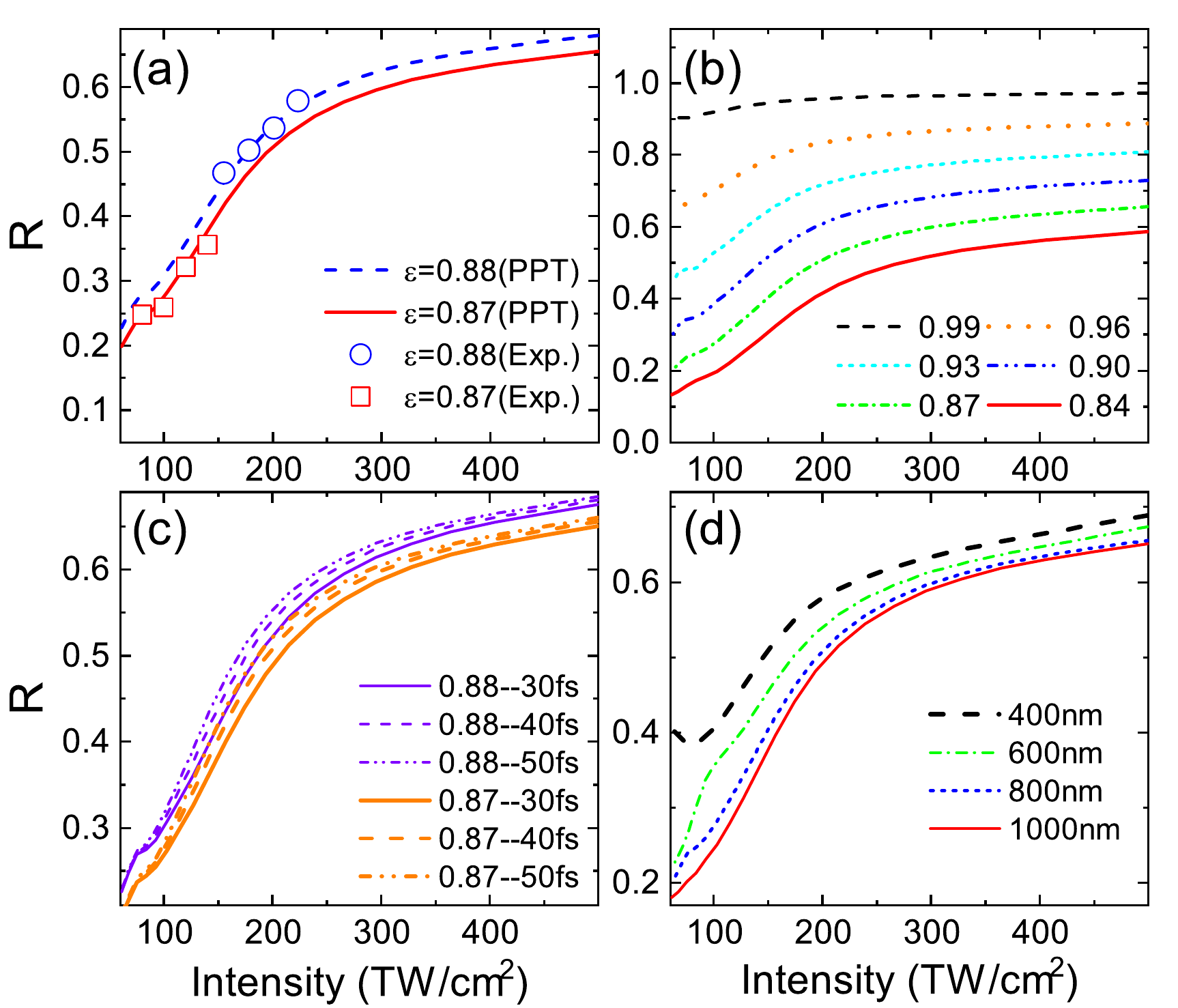}
\end{center}
\caption{(Color online) The modulation $R$   at different laser parameters. (a) and (b) show the dependence of the laser intensity at various $\varepsilon$. Also shown in (a) as symbols are the available experimental data. (c) shows the pulse duration dependence for two different values of $\varepsilon$. (d) shows the wavelength dependence  for $\varepsilon=0.87$. }
\label{fig4}
\end{figure}

 The ion yield oscillation originates from the same physical mechanism with that for the modulation of ATI-spectrum shift. As the ion yield $Y$ is much easier to be measured in most laboratories, we can use the ion yield modulation to extract the ellipticity of the laser pulse. In order to quantify the oscillation of the ion yield $Y$, we  define a modulation index $R = Y_{\rm{min}} /Y_{\rm{max}}$.  At the same laser intensity, one can gradually adjust the ellipticity adopted in the simulation to best fit the curve of $R$ with that of the experimental measurement so that the  $\emph{in situ }$  $\varepsilon$ can be read off. One question is how the laser intensity calibrated experimentally will affect the trend of the modulation index $R$.  { Luckily, it turns out that $R$  only weakly depends on the laser intensity. As examples, we have measured and calculated the modulation at various intensities} with the same ellipticity, three of which are shown in Fig.~3(a) for  $\varepsilon =0.87$. One can see that a variation of 20\% of the intensity only slightly changes the ion yield modulation: the higher the intensity, the shallower the ion modulation.   On the contrary,  $R$ sensitively depends on   $\varepsilon$: the deeper the modulation, the smaller the ellipticity. We only show three cases for such a  strong dependence of $\varepsilon$  in  Fig.~3(b) for the intensity of $200$~TW/cm$^{2}$.  Please note that, in this figure, the value of $\varepsilon$ marked for the experimental data has been evaluated through the traditional optical method.  As one can see,   when the ellipticity is smaller than 0.92, both curves agree with each other rather well, which means that the traditional optical method works reasonably well at smaller $\varepsilon$. However, a large discrepancy is  observed for   $\varepsilon = 0.93$ in which case the present measurement gives a value of  0.95~(i.e., the one used for the calculation to best fit the experimental curve). This deviation further confirms the importance of an $\emph{in situ }$ measurement for a near-circularly polarized pulse.  These comparisons between the simulations and experimental measurements suggest that the modulation   of the ion yield is a good observable for measuring  $\varepsilon$.  In the literature~\cite{yield2}, the ion yields are measured by increasing the pulse energy which can cover a wide range of  intensity. In the present proposal,  by using the counter-rotating  field, the ionization yield can be coherently controlled by accurately varying the electric field within an extremely narrow intensity range in the sub-cycle time scale.

   Now we further investigate theoretically the applicability of the present method in a much   broader range of laser parameters. First of all, let us look at the sensitivity of two ellipticity values of   0.87~(solid line) and 0.88~(dashed line) at a wide range of peak intensity   between 60 to 600 TW/cm$^{2}$, as shown in Fig.~4(a).    The agreement between  the calculations and the available measurements~(symbols) is pretty good, which approves the reliability of the present proposal. Another message Fig.~4(a) conveys is that the modulation is much more sensitive to the change of $\varepsilon$  than the fluctuation of the laser intensities and thus the uncertainty of the intensity calibration will not affect much of the extraction of the ellipticity.  In Fig.~4(b), we extend  the intensity dependence of the modulation to a wide range of $\varepsilon\in[0.84, 0.99]$. One can see that   $R$ is even modulated for $\varepsilon = 0.99$, although the depth is decreased as $\varepsilon$ is increased, especially for higher laser intensities than 200~TW/cm$^{2}$ due to a strong depletion effect.   In Figs. 4(c), as the pulse duration is changed from 30 to 50~fs, the modulation  only has a small difference, which means that the present method can be safely applied in this range  of  pulse durations.  Finally, for pulses with  $\varepsilon = 0.87$ and  the same pulse duration of 40~fs, in Fig.~4(d), we   show  the results for a wide range of the laser wavelength  from 400 to 1000~nm. Generally, the modulation  only weakly depends on the wavelength, especially  in the high-intensity region. For the case of 400~nm,   the possible resonance effect may lead to a larger difference at lower intensities.

In conclusion, we have proposed and realized a robust $\emph{in situ}$  scheme to  retrieve the accurate  value of ellipticity  for a near-circularly polarized laser pulse. We  demonstrated that  the  modulation of the  ion yield  can be served as  a reliable observable for the calibration of $\varepsilon$ in  a wide range of laser parameters. This study   promises a delicate  control and a quantitative interpretation of the sub-cycle  dynamics in  the strong field ionization, the molecular orbital imaging,  and the attosecond dynamics.

 \section{Acknowledgment}

 This work is partially supported by the National Key R\&D Program of China~(Grant No. 2018YFA0306302), and by the National Natural Science Foundation of China (Grants No. 11534004, 11627807, 11725416, 11574010, 11774130) and by the China Postdoctoral Science Foundation( 2015T80293).

\end{document}